\documentclass[epsfig,floats,pre,twocolumn,superscriptaddress,
preprintnumbers,floatfix]{revtex4-1}

\usepackage[latin2]{inputenc}
\usepackage{amsmath}
\usepackage{amssymb}
\usepackage{epsfig}
\usepackage{bm}
\usepackage{color}
\usepackage{stackengine}

\newcommand{\Phio}{\phi_{\circ}}
\newcommand{\Phic}{\phi_\text{c}}
\newcommand{\Phit}{\widetilde{\phi}}
\newcommand{\rc}{r_\text{c}}
\newcommand{\rci}{r_{\text{c},\infty}}
\newcommand{\rco}{r_\text{c,0}}
\newcommand{\rct}{\widetilde{r}_{\text{c}}}
\newcommand{\lm}{\ell_\text{max}}
\newcommand{\tm}{\theta_\text{max}}

\begin{document}

\title{\textbf{Universal Transition to Wide Shear Zones in Entangled Macroscale Chains or Ropes}}

\author{M.\ Reza Shaebani}
\affiliation{Department of Theoretical Physics and Center for Biophysics, 
Saarland University, 66123 Saarbr\"ucken, Germany}
\author{Gerard Gim\'enez-Ribes}
\affiliation{Laboratory of Physics and Physical Chemistry of Foods, Wageningen 
University, 6708WG Wageningen, The Netherlands}
\author{Sybren Zondervan}
\affiliation{Laboratory of Physics and Physical Chemistry of Foods, Wageningen 
University, 6708WG Wageningen, The Netherlands}
\author{Leonard M.\ C.\ Sagis}
\affiliation{Laboratory of Physics and Physical Chemistry of Foods, Wageningen 
University, 6708WG Wageningen, The Netherlands}
\author{Erik van der Linden}
\affiliation{Laboratory of Physics and Physical Chemistry of Foods, Wageningen 
University, 6708WG Wageningen, The Netherlands}
\author{Mehdi Habibi}
\affiliation{Laboratory of Physics and Physical Chemistry of Foods, Wageningen 
University, 6708WG Wageningen, The Netherlands}

\begin{abstract}\noindent
\textbf{Macroscale chains have been proposed to give insight into the physics of molecular 
polymer systems. Nevertheless, understanding the rheological response of systems of 
quasi-one-dimensional semiflexible materials, such as bead-chain packings, is currently 
a great challenge. We study the nonlinear rheology of random assemblies of macroscale 
chains--- including steel bead chains and cooked spaghetti--- under oscillatory shear. 
We show that a universal transition from localized to wide shear zones occurs upon 
increasing the strain amplitude, for a wide range of lengths, flexibilities, and other 
structural parameters of the constituent elements. The critical strain amplitude coincides 
with the onset of strain stiffening development in the system. We obtain scaling laws 
for transition sharpness, shear-zone width, and stiffness enhancement as a function of 
chain length. Our findings suggest that the entanglements between the constituent elements 
strengthen when approaching the critical strain amplitude and rapidly become long range, 
even spanning the entire finite system for long enough chains. We show that the nonlinear 
rheological response is governed by the interplay between increasing stored elastic forces 
due to entanglements and increasing contribution of dissipation with shear rate and 
interlocking between chains.}
\end{abstract}
 
\maketitle

\noindent 
Disordered assemblies of long semiflexible objects are a class of materials ubiquitously 
observed in nature and daily life. Examples are bird nests \cite{Hansell05}, aegagropila 
networks \cite{Verhille17}, and unwoven textiles \cite{Weiner20}. An intriguing and 
important property of such entangled assemblies is their mechanical response and yielding 
when subject to external stresses. For example, packings of bead chains exhibit a striking 
strain stiffening under shear \cite{Brown12,Dumont18}. The complexity arises from the 
presence of topological constraints such as (semi)loops, knots, and interlocking between 
the constituent elements \cite{Zou09,Gomez20,Soh19,BenNaim01,Lopatina11,Dumont18,Sarate22}. 
The behavior differs considerably from that of packings of rodlike objects, where the 
physics is mainly governed by frictional interactions, volume exclusion, and aspect 
ratio of rods. Despite the analogies with polymeric materials \cite{Safford09,Taheri18}, 
developing a quantitative theory of stability and unjamming response of macroscale 
chain assemblies requires a detailed understanding of the roles of entanglements and 
friction, which is still lacking. The insight from the mechanical response of 
entangled-driven athermal systems of long semiflexible objects can guide future 
design of new smart unwoven textiles \cite{Sunami22,Hu12,Yun13}, knitted fabrics 
\cite{Poincloux18}, and artificial mussels and tissues \cite{Haines16}.

Upon yielding, slowly sheared packings of individual beads often form rigid regions 
separated by narrow shear zones near moving boundaries where the material flows with 
a shear-rate-independent profile shape \cite{Mueth00,Losert00} (though wide shear 
zones have also been reported when shearing the bulk material away from the boundaries 
\cite{Shaebani21,Fenistein03}). While shear banding in frictional granular materials 
can be understood based on energy dissipation considerations \cite{Unger04,Moosavi13}, 
much less is known about the yielding of semiflexible chain assemblies, particularly, 
whether and how far entanglements and chain length and flexibility broaden the shear 
zones and extend them into the bulk. Since shear zones mark regions of material failure 
and energy dissipation, understanding the yielding behavior of chain assemblies is 
crucial in industrial processes and for design of new disordered meta materials 
\cite{Aktas21,Weiner20,Verhille17,Mirzaali17}. As a daily-life application, by twisting 
cooked spaghetti on a plate with a fork, an interesting question is how the amount 
of rolled spaghetti around the fork depends on length, softness, and adhesion of 
the strands?

The rheological response of viscoelastic materials is of fundamental importance in 
physics, engineering, and biology. There has been growing interest in nonlinear 
viscoelastic responses to large strains \cite{Kamani21,MatozFernandez17,Rogers18,
Baggioli20}, e.g., to differentiate between materials with similar linear but 
drastically different nonlinear responses.--- A highly informative protocol is to 
apply an oscillatory shear strain \cite{Rogers18,Baggioli20,Tapadia06}: After 
many cycles to become independent of the prior history of the sample, the 
steady-state stress response can be probed over a wide strain range from below 
to above the yield point.--- Nevertheless, the rheology of macroscale chain 
assemblies remains less explored \cite{Regev13}. It is unclear how the interplay 
between topological constraints and dissipation governs the nonlinear rheological 
physics of these systems.

In the present work, we study the rheological response of bead-chain assemblies 
to oscillatory shear deformations in experiments and numerical simulations and 
compare it to that of cooked spaghetti. We observe a striking universal transition 
from narrow shear zones at small amplitude oscillatory shear (SAOS) to wide shear 
zones at large amplitude oscillatory shear (LAOS). Increasing the length of the 
constituent elements sharpens the transition and enhances the extent of the wide 
shear zone, for which power-law scaling relations are obtained. Our results show 
that the system undergoes a rather sharp crossover from inactive entanglements in 
SAOS to system-spanning activated entanglements in LAOS. We demonstrate that the 
nonlinear rheological physics of macroscale chain assemblies is governed by the 
competition between stored elastic and dissipative (viscous) forces: The applied 
shear strain enhances the elastic contribution to stress by strengthening the 
topological constraints while the contribution of dissipation--- which is 
proportional to the shear rate--- grows above the yield point and also with 
increasing the interlocking between the constituent elements. 

\begin{figure}[t]
\centerline{\includegraphics[width=0.44\textwidth]{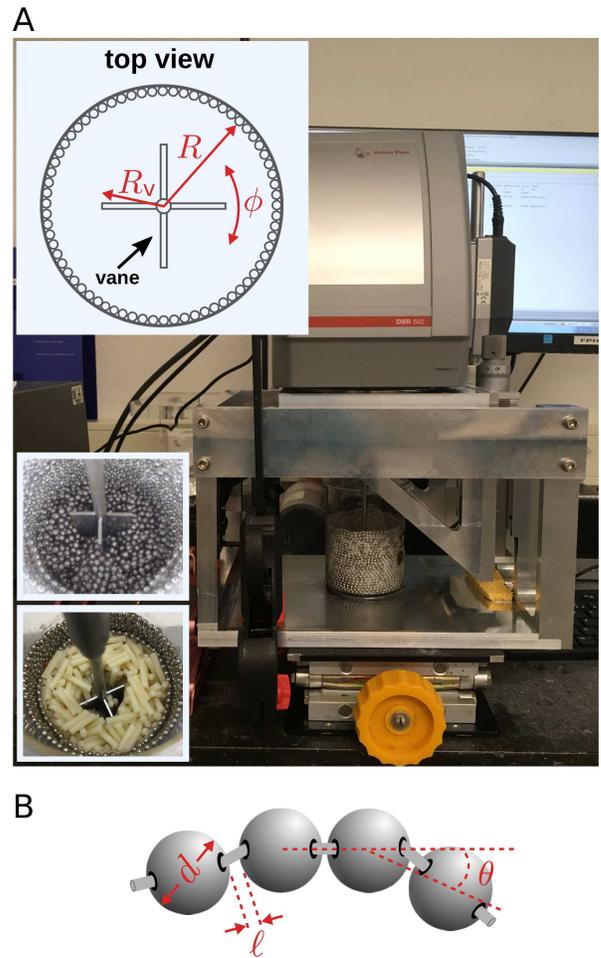}}
\caption{\textbf{Setup and bead-chain geometries.} (A) Four-blade vane rheometer setup. 
Upper inset: top view of the cell, not to scale. $R_\text{v}\,{=}\,11\,\text{mm}$; $R\,
{\simeq}\,34\,\text{mm}$. Lower insets: images of shear experiments with steel bead 
chains or cooked spaghetti. (B) Schematic of the bead chain with $\ell$ and $\theta$ 
indicating the bond length and turning angle, respectively, and $d\,{=}\,2.4\,\text{mm}$ 
being the bead diameter. $\ell$ is variable from $0$ to $\lm\,{=}\,0.4\,\text{mm}$ and 
$\theta$ from $0$ to $\tm\,{\simeq}\,40^{\circ}$ in experiments.}
\label{Fig:1}
\end{figure}

Our rheometer setup shown in Fig.\,\ref{Fig:1}A consists of a cylindrical container 
of inner radius $R$ and a rotating four-blade vane with blade radius $R_\text{v}$, 
which applies a sinusoidal deformation 
\begin{equation}
\phi\,{=}\,\Phio \sin(2\pi f t),  
\end{equation}
where $\phi$ is the deflection angle, and $\Phio$ and $f{=}\,0.1\,\text{Hz}$ denote 
the rotation amplitude and frequency, respectively. As a model material, we use 
either cooked spaghetti (which is cut into equal pieces) or bead chains consisting 
of hollow spherical beads with diameter $d$ flexibly connected to each other by 
enclosing dog-bone-shaped links. The induced gap size $\ell$ between the neighboring 
beads varies in the range $0{\leq}\ell{\leq}\ell_\text{max}$. The links also limit 
the local turning angle $\theta$ of the chain to $0{\leq}\theta{\leq}\tm$; see 
Fig.\,\ref{Fig:1}B and \emph{Materials and Methods} section for details. An 
instantaneous persistence $p{=}\text{cos}(\theta)$ can be assigned to the chain 
nodes \cite{Shaebani20} from which the local persistence length $\ell_p$ can be 
obtained via $p{=}\text{e}^{-\ell{/}\ell_p}$ \cite{Doi86,Taheri18,Mortazavi16}. 
We also perform extensive contact dynamics (CD) simulations \cite{Shojaaee12} 
of spherical rigid beads in a setup similar to our experiments. We impose upper 
bounds on the distance between the centers of neighboring beads and on the angle 
between lines connecting three neighboring beads along the chain. This concept 
is suited very well to the CD method where interparticle forces are handled as 
constraint forces \cite{Jean99,Shaebani09}.  
\smallskip

\noindent\textbf{Universal transition to wide shear zones} 
\smallskip

\begin{figure*}[t]
\centerline{\includegraphics[width=0.95\textwidth]{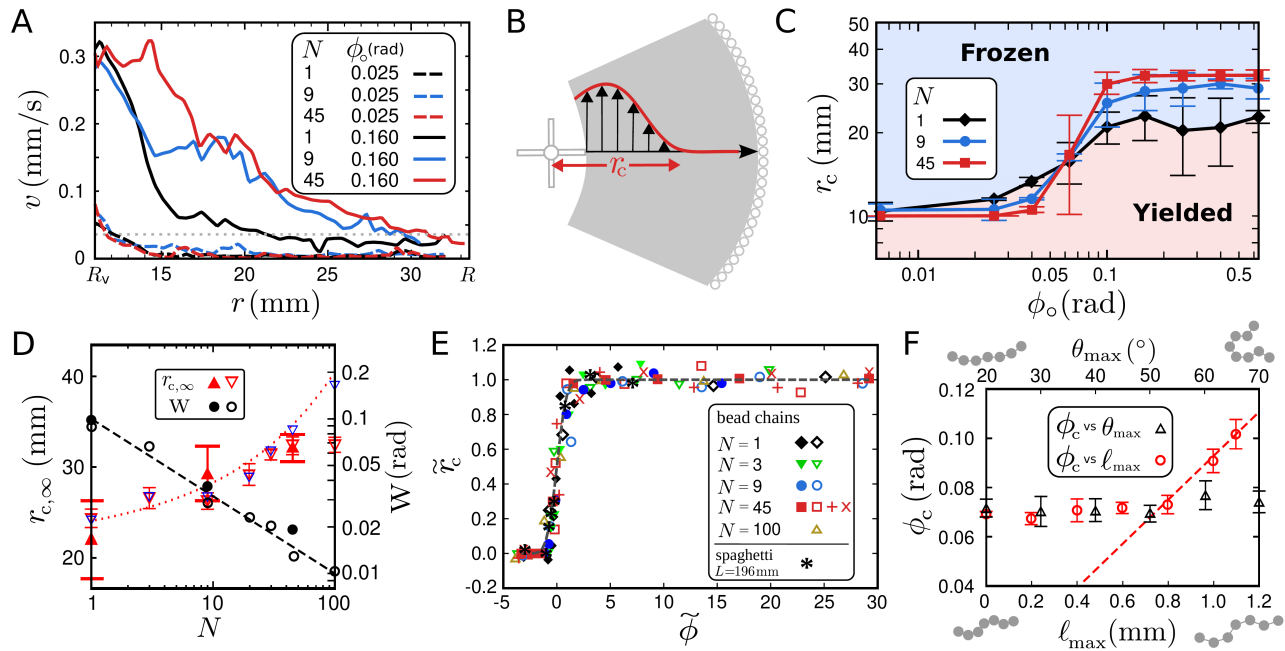}}
\caption{\textbf{Chain dynamics in oscillatory shear.} (A) Mean bead velocity $v$ as 
a function of distance $r$ from the cylinder axis for different values of rotation 
amplitude $\Phio$ and chain length $N$. The dotted line denotes the threshold velocity 
$v_\text{c}$. (B) Definition of the shear-zone width $\rc$. The velocity field is 
sketched with vertical arrows. (C) Shear-zone width $\rc$ versus rotation amplitude 
$\Phio$. The yielded and frozen regions for $N{=}1$ are indicated with different colors. 
(D) Saturation width $\rci$ of the shear zone and transition sharpness $W$ (indicating 
the range of rotation amplitudes over which the transition occurs) in terms of $N$. 
Full (open) symbols correspond to experimental (numerical) results. Blue open symbols 
represent the simulation results for a larger container with $R\,{\simeq}\,100\,
\text{mm}$. The black dashed line is a fit to Eq.\,(\ref{Eq:WN}), and the red dotted 
line represents a growth according to Eq.\,(\ref{Eq:rciN}), as a guide to the eye. 
(E) Collapse of all rescaled shear-zone widths $\rct\,{=}\,\frac{\rc\,{-}\,\rco}{\rci\,
{-}\,\rco}$ as a function of rescaled rotation amplitude $\Phit\,{=}\,\frac{\Phio\,
{-}\,\Phic}{W}$, obtained for different chain lengths in experiments (full symbols) 
and simulations (open symbols). The simulation results for a chain with zero bond 
length $\lm\,{=}\,0$ (pluses) or high flexibility $\tm\,{=}\,70^{\circ}$ (crosses) 
at $N{=}45$ are also presented. The dashed line is a fit to the error function 
(\ref{Eq:Erf}). (F) Transition center $\Phic$ versus maximum bond length $\lm$ 
or maximum turning angle $\tm$ obtained from simulations for $N{=}45$. The line 
represents $\Phic\,{=}\,\lm{/}R_\text{v}$.}
\label{Fig:2}
\end{figure*}

\noindent We start the experiments in the SAOS regime, i.e.\ with small rotation amplitudes. 
In this regime, the surface velocity profiles reveal that the movements diminish rapidly 
with increasing distance from the rotating blades, independent of the chain length $N$; 
see e.g.\ the experiments at $\Phio\,{=}\,0.063\,\text{rad}$ in \emph{Suppl.\ Movie S1}. 
Denoting the mean velocity at the radial coordinate $r$ with $v(r)$, Fig.\,\ref{Fig:2}A 
shows that $v$ decays fast with $r$; a jammed immobile region is reached after 1-2 bead 
diameter distance. The velocity fluctuations are relatively small and the profiles are 
reproducible to a large extent independent of the initial conditions. Formation of narrow 
localized shear zones near moving boundaries was observed in quasistatic shear of granular 
materials \cite{Mueth00,Losert00}. Here, the independence of the results from $N$ indicates 
that the topological constraints have not been activated yet and do not play a role in the 
SAOS regime. 

To quantify the extent of the shear zone, we assign a width $\rc$ to the shear zone as the 
radial distance from the cylinder axis at which the velocity drops below a threshold value; 
see Fig.\,\ref{Fig:2}B. Here we report the results for the threshold velocity $v_\text{c}\,
{=}\,4{\times}10^{-5}\,\text{m/s}$; however, we checked that the observed trends and our 
conclusions are insensitive to this choice in a moderate range of velocities. 

The behavior is markedly different in the LAOS regime: The shear zone extends to the 
bulk of the system, as shown in Fig.\,\ref{Fig:2}A (see also \emph{Suppl.\ Movie S1} 
for experiments at $\Phio\,{=}\,0.632\,\text{rad}$). It was previously reported that 
oscillatory shear of granular materials with large amplitudes broadens the shear zone 
up to a few bead diameters, as the shear stress drops at the shear direction reversals 
\cite{Toiya04}. Our notable observation is the influence of chain length on the shear 
deformation at large strains: With increasing $N$, the shear zone becomes much wider, 
the velocity profile strongly depends on the initial condition, and large velocity 
fluctuations along the radial coordinate are observed even for a single experiment. 

To understand how the system crosses over from narrow shear zones in SAOS to wide 
system-spanning ones in LAOS, we vary systematically $\Phio$ for different chain 
lengths and repeat the measurement for different non-consecutive cycles in each 
sample to smoothen the velocity fields. The width $\rc$ of the shear zone as a 
function of $\Phio$ is shown in Fig.\,\ref{Fig:2}C. Interestingly, the change 
of $\rc$ when moving from the SAOS to the LAOS regime is not gradual but happens 
over a narrow range $W$ of rotation amplitudes. For longer chains the transition 
is sharper, corresponding to a smaller $W$. By fitting each $\rc{-}\Phio$ curve 
to an error function, we assign a $W$ to it. The plot of the resulting $W$ values 
in terms of $N$ in Fig.\,\ref{Fig:2}D implies that $W$ scales as
\begin{equation}
W \sim N^{-\alpha},
\label{Eq:WN}
\end{equation}
with the exponent $\alpha\,{\simeq}\,0.5$. 

\begin{figure*}[t]
\centerline{\includegraphics[width=0.95\textwidth]{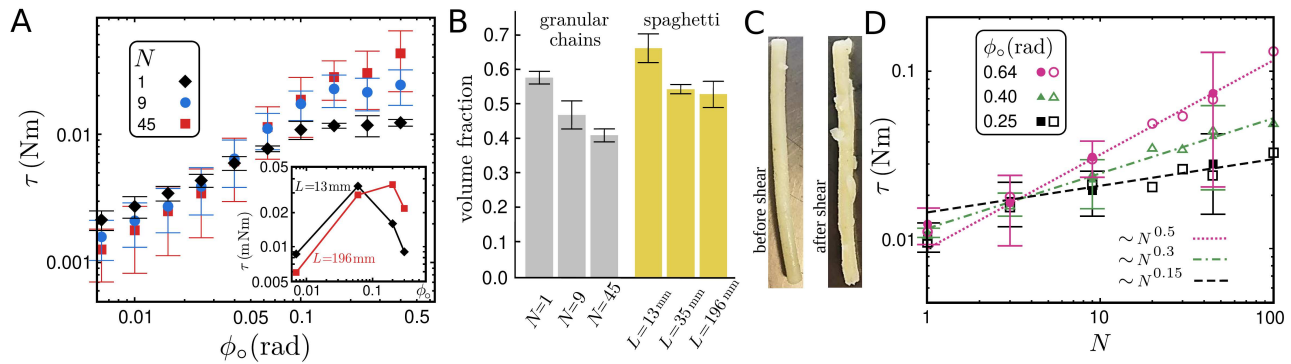}}
\caption{\textbf{Stiffening of macroscale chains.} (A) Torque $\tau$ versus rotation 
amplitude $\Phio$ for different bead chain lengths. Inset: $\tau$ vs $\Phio$ in shear 
experiments with cooked spaghetti of length $L$. (B) Volume fraction of bead chain or 
spaghetti assemblies. (C) Example of surface roughening of spaghetti due to oscillatory 
shear. (D) $\tau$ versus $N$ for different values of $\Phio$ in experiments (full 
symbols) and simulations (open symbols). The lines represent power-law fits to 
$N{\geq}3$ data.}
\label{Fig:3}
\end{figure*}

Figure\,\ref{Fig:2}C also reveals that $\rc$ reaches a plateau level at large $\Phio$, 
which is higher for longer chains. Denoting this maximum shear-zone width with $\rci$, 
Fig.\,\ref{Fig:2}D shows that $\rci$ initially grows with $N$ but gradually approaches 
the inner radius $R$ of the container (full triangles). The saturation behavior at 
larger $N$ is more visible in the simulation results (red open triangles). The question 
arises whether the shear zone for long chains can penetrate further into the bulk of 
the system in an infinite system (i.e.\ when $R{\rightarrow}\infty$). To answer this, 
we perform simulations with chain lengths up to $N{=}100$ and increase the container 
radius from $R\,{\simeq}\,34$ to $67$ and $100\,\text{mm}$. As the resulting $\rci$ 
values for the two latter system sizes are the same (within the error bars), we 
conclude that the shear zones for chain lengths $1{\leq}N{\leq}100$ cannot be wider 
if the system size is further increased beyond $R\,{=}\,100\,\text{mm}$. A comparison 
between the results at $R\,{=}\,34$ (open red triangles) and $R\,{=}\,100\,\text{mm}$ 
(open blue triangles) in Fig.\,\ref{Fig:2}D shows the finite-size effects on 
$\rci$ values for longer chains. The system-size-independent width of the shear 
zone (blue symbols) can be approximately described by
\begin{equation}
\rci(N)\,{-}\,\rci(N{=}1) \sim N^{\beta},
\label{Eq:rciN}
\end{equation}
with $\beta\,{\approx}\,0.5$. The radius of gyration of a flexible chain similarly 
scales with $\sqrt{N}$ \cite{Doi86}. This suggests that in the LAOS regime the 
topological constraints are fully activated along the entire length of the chains 
that are moved by the rotating blades. As a result, their motion can influence a bulk 
region of size proportional to their radius of gyration, i.e., ${\propto}\,\sqrt{N}$.

Figure\,\ref{Fig:2}E illustrates our main result: After proper rescaling, all $\rc{-}
\Phio$ data collapse on a universal curve which is well fitted by an error function. 
We introduce the rescaled shear-zone width $\displaystyle\rct\,{=}\,\frac{\rc\,{-}\,
\rco}{\rci\,{-}\,\rco}$, with $\rco$ being the minimum shear-zone width. Thus, the 
dimensionless number $\rct$ takes values in $[0,1]$. All rescaled shear-zone widths  
follow a universal master curve
\begin{equation}
\rct=\displaystyle\frac12 \Big(1\,{+}\,\text{erf}\big(\frac{\Phio\,{-}\,
\Phic}{\text{W}}\big)\Big),
\label{Eq:Erf}
\end{equation}
where $W$ is the range of amplitudes over which the transition occurs (transition 
width) and $\Phic$ is the rotation amplitude at the center of the error function 
(transition center). We checked that the spaghetti data and also simulation results 
for $1{\leq}\,N\,{\leq}100$ and other values of the maximum bond length in the 
range $0\,{\leq}\,\lm\,{<}\,d{/}2$ or maximum turning angle in the range $20^{\circ}
\,{\leq}\,\tm\,{<}\,70^{\circ}$ are well fitted to Eq.\,(\ref{Eq:Erf}).
  
From the fits of $\rc{-}\Phio$ curves to an error function, we find that the transition 
center $\Phic$ does not depend on $N$. This can be also seen from Fig.\,\ref{Fig:2}C; 
we get $\Phic\,{\simeq}\,0.07\,\text{rad}$ by averaging over experiments with different 
$N$. Simulations with different values of $\tm$ or $\lm$ shown in Fig.\,\ref{Fig:2}F 
reveal that $\Phic$ is independent of the chain flexibility $\tm$. Moreover, $\Phic$ 
is insensitive to variation of $\lm$ for small bond lengths but grows approximately 
linearly for $\lm\,{\gtrsim}\,0.8\,\text{mm}$. 

A plausible scenario is that the transition center $\Phic$ is the onset at which the 
shearing brings the bonds between neighboring beads to their maximum possible length 
$\lm$ quickly after each shear direction reversal. Above this threshold rotation 
amplitude, the chains are in a stretched form in the absence of the internal degrees 
of freedom of having variable bond lengths. This would lead to the prediction $\Phic
\,{\approx}\,\frac{\lm}{R_\text{v}}$, i.e.\ a linear increase of $\Phic$ with $\lm$. 
Figure\,\ref{Fig:2}F shows that this simple model captures the behavior for long bonds. 
But why does not $\Phic$ grow proportionally to $\lm$ for short bonds? From simulations 
with $N{=}1$, we observe that a minimum rotation amplitude of the order of $0.06{-}0.07
\,\text{rad}$ is required to generate wide shear zones in packings of individual beads. 
That is why the increase of $\Phic$ with $\lm$ is only visible for bond lengths $\lm\,
{\gtrsim}\,0.8\,\text{mm}$ for which $\Phic\,{\gtrsim}\,0.07\,\text{rad}$. 
\smallskip

\noindent\textbf{Stiffening upon increasing strain amplitude} 
\smallskip

\noindent The crossover from narrow to wide shear zones suggests the presence of 
entanglements above the transition threshold $\Phic$. One of the entanglement mechanisms 
is the interlocking between the beads of different chains. We expect that stretching of 
bonds to $\lm$ at $\Phic$ immediately activates this type of topological constraints for 
all chain lengths $N{\geq}3$. Formation of (semi)loops is another entanglement mechanism, 
which strengthens with increasing $N$ and/or $\Phio$ above $\Phic$. Note that a full 
ring requires a minimum chain length $N{=}9$. The fact that the shear-zone width rapidly 
saturates above $\Phic$ for all $N$ shows that interlocking is the major entanglement 
mechanism affecting the flow properties of chain assemblies. Loop formation induces weak 
entanglements for $N{<}9$ and, moreover, should lead to $\Phio$-dependent wide shear zones 
which is not observed. It is however expected that semiloops play the major role in the 
strain stiffening phenomenon \cite{Brown12,Dumont18}. 

\begin{figure}[t]
\centerline{\includegraphics[width=0.45\textwidth]{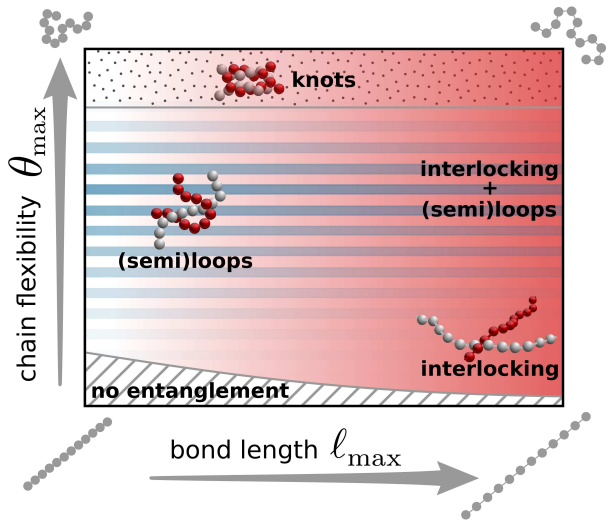}}
\caption{\textbf{Schematic phase diagram of entanglement mechanisms at a given chain length.} 
The contribution of (semi)loops grows with increasing chain flexibility from rod-like chains; 
however, extremely flexible chains promote knot formation. Increasing bond length gradually 
enhances the contribution of interlocking between chains. Gradients of red and blue colors 
qualitatively indicate variations in the contributions of interlocking and semiloops, 
respectively.}
\label{Fig:4}
\end{figure}

As a direct proof of the activation of entanglements, we probe the mechanical response 
of the system upon increasing $\Phio$. By measuring the maximum exerted torque $\tau$ 
by the rheometer on the system in an oscillatory shear cycle for a given rotation amplitude 
$\Phio$ and repeating it for different $\Phio$ values, we plot $\tau$ as a function of $\Phio$ 
in Fig.\,\ref{Fig:3}A. In the absence of entanglements at small strains, shearing the packings 
of individual beads ($N{=}1$) requires a larger $\tau$ compared to long-chain packings, due 
to having a higher packing fraction; see Fig.\,\ref{Fig:3}B and Ref.\,\cite{Zou09} (Although 
packing structure strongly depends on the interparticle friction \cite{Shaebani08,Unger05,
Shaebani09b}, we assume that it affects the assemblies with different $N$ in a similar way). 
In contrast, at large amplitude oscillations, yielding occurs for $N{=}1$ while assemblies 
of longer chains exhibit shear stiffening. The crossover point in Fig.\,\ref{Fig:3}A 
interestingly coincides with the onset $\Phic$ of the transition to wide shear zones. We 
observe a similar transition point and mechanical response trend in the case of spaghetti 
(inset of Fig.\,\ref{Fig:3}A), even though they cannot sustain large torques in the strain 
stiffening regime: Large amplitude shear of spaghetti leads to roughening of strand surfaces 
(Fig.\,\ref{Fig:3}C), cutting of strands into shorter pieces, and ascending and ordering 
of strands above the probe. Moreover, it was previously reported that shear rate (which 
increases with $\Phio$ in our experiments) and interfilament overlap length enhance the 
sliding friction between filaments \cite{Ward15}. Nevertheless, observing a similar onset 
of stiffening for bead chains and spaghetti is striking and it possibly points to a universal 
mechanical response of assemblies of long semiflexible objects. 

\begin{figure}[t]
\centerline{\includegraphics[width=0.47\textwidth]{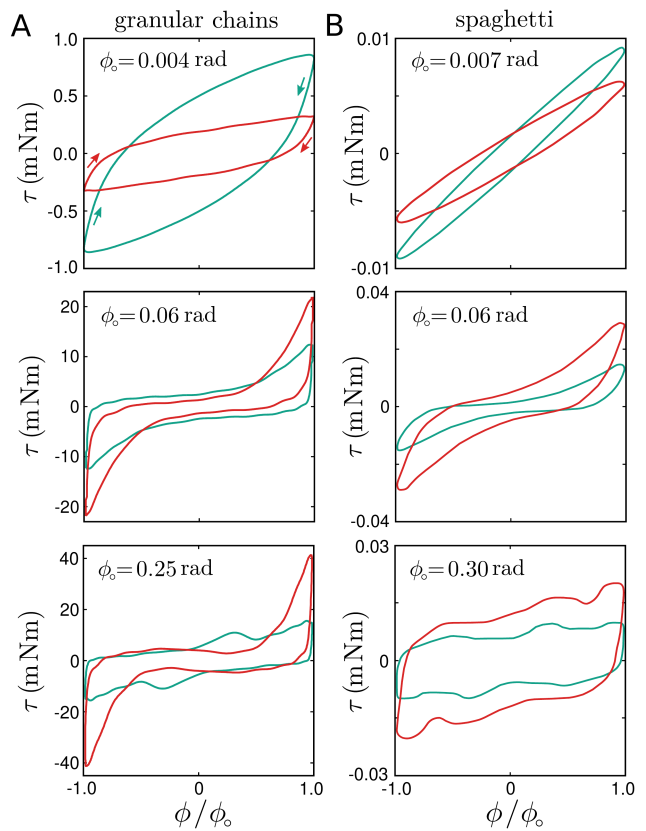}}
\caption{\textbf{Elastic Lissajous curves of sheared bead chains and spaghetti.} 
(A) Lissajous plots of torque vs deflection angle $\phi$ (scaled by $\Phio$) 
for bead chains of length $N{=}1$ (green) and $N{=}45$ (red). (B) Similar 
plots for assemblies of spaghetti with length $L{=}13\,\text{mm}$ (green) 
and $L{=}196\,\text{mm}$ (red).}
\label{Fig:5}
\end{figure}

Figure\,\ref{Fig:3}A shows that the degree of stiffening not only depends on $\Phio$ but also 
on $N$ \cite{Brown12}. When plotting $\tau$ vs $N$ for different $\Phio$ in Fig.\,\ref{Fig:3}D, 
we find that $\tau$ scales as a power-law with the chain length (for $N{\geq}3$)
\begin{equation}
\tau \sim N^{\gamma(\Phio)}.
\label{Eq:TauN}
\end{equation}
The exponent $\gamma$ depends on $\Phio$; it increases from $\gamma\,{=}\,0$ at $\Phio{=}
\Phic$ and reaches, e.g., $\gamma\,{\simeq}\,0.5$ at $\Phio\,{\simeq}\,0.64\,\text{rad}$. 
By varying $\lm$ in simulations from $0$ to ${\sim}d{/}3$ to increase the contribution 
of interlocking, we find a nearly $20\%$ increase in $\tau$, reflecting the greater 
contribution of semiloops to the mechanical response. For a given chain length, 
Fig.\,\ref{Fig:4} summarizes the effective entanglement mechanisms in the ($\lm$, $\tm$) 
space, i.e.\ upon varying chain flexibility and interparticle bond length.  
\smallskip

\begin{figure}[t]
\centerline{\includegraphics[width=0.4\textwidth]{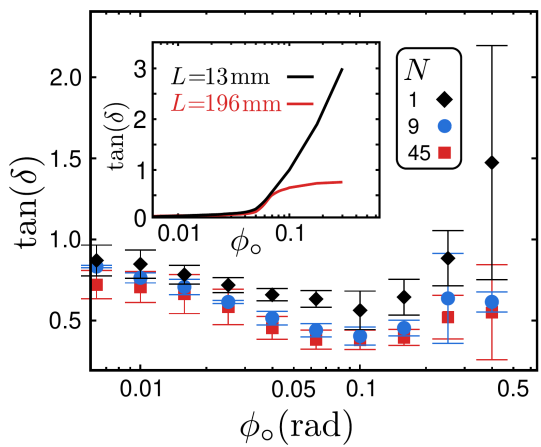}}
\caption{\textbf{Chain-length dependence of the loss tangent.} The main panel 
represents the loss tangent $\text{tan}(\delta)$ versus rotation amplitude 
$\Phio$ for different bead chain lengths. The inset represents a similar 
comparison between assemblies of spaghetti with different lengths.}
\label{Fig:6}
\end{figure}

\noindent\textbf{Nonlinear rheological response} 
\smallskip

\noindent To clarify the similarities and differences between the rheological response 
of bead-chain and spaghetti assemblies, we apply many cycles of oscillatory shear to 
reach the steady-state stress response. Next, we plot the torque $\tau$ and the deflection 
angle $\phi$ in one shear cycle in the form of parametric Lissajous-Bowditch curves, 
similar to the stress vs strain curves commonly used in the literature \cite{Ewoldt08} 
(As the calculation of stress and strain in our inhomogeneously sheared system involves 
approximations, here we use raw $\tau$-$\phi$ data to avoid approximation errors). In 
the SAOS regime (Fig.\,\ref{Fig:5}, top row), we observe that the spaghetti data display 
an ellipsoidal shape, characteristic of a linear viscoelastic response. The response 
of granular chains displays a rhomboidal shape. The nonlinear behavior indicates that 
higher harmonics are present in the signal. The degree of stiffening is larger for 
shorter constituent elements (being bead chains or spaghetti strands) in agreement 
with Fig.\,\ref{Fig:3}A. Around the transition amplitude $\Phic$ (middle row), a 
highly nonlinear torque response develops in both systems, evidencing the emergence 
of intracycle stiffening. Here, the curves develop a larger degree of stiffening 
for assemblies of longer constituent elements. We also note that the stiffening in 
bead-chain Lissajous curves is more pronounced due to the simultaneous formation 
of loops and activation of interlocking; the latter is absent in spaghetti assemblies. 
The main difference between the rheological response of bead-chain and spaghetti 
assemblies arises in the LAOS regime (bottom row): In bead chains, while packings 
of individual beads ($N{=}1$) yield, assemblies of longer chains exhibit an even 
stronger response. In contrast, spaghetti assemblies are unable to develop higher 
stress and yield at all strand lengths; still, the response remains steeper for 
longer spaghetti strands.

Finally, we clarify how viscous stresses develop with increasing the rotation amplitude 
$\Phio$ in macroscale chain assemblies. This can be achieved, e.g., by analyzing viscous 
Lissajous curves of $\tau$ vs $\dot{\phi}$. Here, we alternatively consider the loss 
tangent, $\text{tan}(\delta)$, which is a dimensionless parameter that measures the 
ratio of dissipated to stored energy in one cycle of oscillation. The results shown 
in Fig.\,\ref{Fig:6} reveal that packings of individual beads dissipate a larger 
fraction of energy compared to chain assemblies, even in SAOS. The difference becomes 
more pronounced in LAOS due to the frictional flow of the yielded assemblies of 
individual beads as well as the increased contribution of the stored elastic energy 
in chain packings upon strengthening the entanglements. A similar trend is observed 
in shearing of spaghetti, i.e., longer strands dissipate less energy. Here, the role 
of the stored elastic energy is less important since longer strands fail to develop 
entanglements at large shear strains. Instead, the frictional flow of shorter strands 
is more dissipative. 
\smallskip

\noindent\textbf{Conclusion} 
\smallskip

\noindent In conclusion, we have studied the nonlinear rheology of entangled assemblies 
of long semiflexible objects. Unveiling the rheological response of externally-driven 
interacting particle assemblies--- particularly quasi-1D materials such as (bio)polymers 
and granular chains--- is currently a great challenge for statistical physics. Our results 
demonstrate a universal transition from narrow shear zones at low amplitudes of oscillatory 
shear to broad ones at large amplitudes with a width that scales with the chain length. 
We have linked this intriguing rheological response to the development of topological and 
geometrical constraints: The system undergoes a sharp crossover from lacking entanglements 
to a highly entangled environment. Nevertheless, entanglements are not the only influential 
factor in determining the mechanical response and yielding of macroscale chain assemblies. 
Friction is known to play a crucial role in packings of individual beads and networks of 
fibres \cite{Goldenberg05,Shaebani07,Negi21}. Understanding how the interplay of friction 
and entanglements governs the rheological response of chain assemblies is a future challenge 
toward answering the question of how assemblies of long flexible objects flow. The insight 
from the present study can also help for better understanding the compaction and packaging 
of quasi-1D semiflexible objects \cite{Stoop11,Shaebani17,Vetter14} and can be used as a 
guide to design the micro/nano-structure of new materials.
\smallskip

\noindent\textbf{Materials and methods} 
\smallskip

\noindent\footnotesize{\textbf{Vane rheometer}
\smallskip

\noindent A rheometer measuring head (Anton-Paar DSR 502, Austria, Graz) was mounted 
on a vertically moving frame with a digimatic indicator (ID-H Series, Mitutoyo, The 
Netherlands, Veenendaal) to accurately adjust the height of the probe. A custom made 
cup, consisting of a 250 mL glass beaker roughened with bead chains, was attached to 
the bottom plate. A four-blade vane (ST22-4V-40, Serial No.: 18180, Anton-Paar, Austria, 
Graz) of 22 mm diameter and 40 mm height was used. To record the movements at the top 
surface, a GoPro Hero 4 Silver camera (GoPro, U.S., San Mateo) was used with a frame 
rate of 25 $\text{frames}{/}\text{s}$. The probe was first inserted into the cup and 
secured by hand while loading the sample. To avoid exceeding the maximum torque that 
the rheometer could apply, the bead chains or spaghetti were loaded into the cup up 
to 2 cm and 4 cm height, respectively. After loading the spaghetti samples, they were 
gently pushed downwards without being fractured. For experiments with spaghetti a lid, 
made of a low-density polyethylene petri dish and secured between an additional layer 
of chains on the beaker wall, was used to prevent the samples from ascending above 
the probe. In case of bead chains, the beaker was vibrated for 2 mins on a Vortex-Genie 
2 test tube shaker (Scientific Industries, U.S., Bohemia) to reach the maximum possible 
packing fraction. Next, the probe was connected to the rheometer head and pre-sheared 
at 1 Hz frequency and 63 mrad rotation amplitude for 4 mins. Finally, the oscillatory 
shear tests were carried out at 0.1 Hz frequency and over a rotation amplitude range 
limited by the maximum torque that the rheometer could apply ($\Phio{\leq}0.6\,\text{rad}$ 
for bead chains and $\Phio{\leq}1.9\,\text{rad}$ for spaghetti). 
\smallskip

\noindent\footnotesize{\textbf{Bead chains}
\smallskip

\noindent Stainless steel granular bead chains (Grootspul, The Netherlands, Driebergen) 
were used with a bead diameter of 2.4 mm, a maximum distance of 2.8 mm between the centers 
of neighboring beads, and a maximum turning angle of $40^{\circ}$, resulting in a minimum 
loop circumference of 9 beads. The bead chains were cut into lengths of 3, 9, and 45 beads, 
as well as monomers. Sunflower oil (Vandemoortele Nederland BV, The Netherlands, Zeewolde) 
was used as a lubricant to reduce the friction coefficient to $\mu\,{\approx}\,0.2$; the 
chains were coated by shaking them for 10 mins in plastic containers using 0.02 mL of oil 
per gram of beads.
\smallskip

\noindent\footnotesize{\textbf{Cooked spaghetti}
\smallskip

\noindent Dehydrated dry spaghetti (Jumbo Spaghetti Naturel, The Netherlands, Veghel) were 
cut into lengths of 1.0, 3.0, 9.0 and 15.0 cm. Excess starch was cleaned by submerging the 
spaghetti in cold tap water for 10 s twice, while gently stirring using a spatula. The 
samples were subsequently boiled for 9 mins and cooled in a sieve using running cold tap
water. The majority of the water was drained. Next, sunflower oil (Vandemoortele Nederland
BV, The Netherlands, Zeewolde) was used as a lubricant; the samples were coated using 0.02 
mL oil per gram of boiled spaghetti by stirring and subsequent shaking by hand for 5 mins 
in a closed plastic container. Due to swelling of the spaghetti during cooking, the final 
strand lengths were $1.3\,{\pm}\,0.2$, $3.5\,{\pm}\,0.1$, $12.3\,{\pm}\,0.2$ and $19.6\,
{\pm}\,0.2$ cm. The spaghetti samples were stored in closed containers at room temperature 
and measured within eight hours after preparation.
\smallskip

\noindent\footnotesize{\textbf{Image processing}
\smallskip

\noindent Particle image velocimetry (PIV) was performed using PIVlab 2.55 plugin \cite{Thielicke14}, 
a tool developed for MATLAB R2019b (MathWorks, U.S., Natick). The settings for PIVlab were 
used as described in \cite{Sarno19}. Four passes were applied for PIV estimation with the 
first being approximately four times and the last about a half of the bead diameter. The 
images were calibrated for time and space using the time between successive frames and 
the width of the beaker at the sample surface, respectively. 
\smallskip

\noindent\textbf{Simulation method} 
\smallskip

\noindent Simulations of bead chains were performed using the contact dynamics (CD) method 
\cite{Shojaaee12,Jean99}. To construct the chains, spherical rigid beads of diameter $d$ 
were connected to each other by imposing an upper bound on the distance $d{+}\ell$ between 
the centers of neighboring beads, with $\ell$ being the gap between the surfaces of the two 
beads. The maximum gap size $\lm$ varied within $0\,{\leq}\,\lm\,{<}\,d{/}2$ in different 
simulations; thus, the gap between beads on the same chain was always smaller than the bead 
radius to ensure that the neighboring beads remain so close to each other that the imaginary 
bonds between the beads of different chains never touch. To tune the flexibility of the chains, 
the angle $\theta$ between the lines connecting the centers of successive bead pairs on the 
chain was limited to an upper bound $\tm$ ($\tm\,{\in}\,\{20^{\circ},30^{\circ},40^{\circ},
50^{\circ},60^{\circ},70^{\circ}\}$ in different simulations). The bead chain length was 
varied from $N{=}1$ (individual beads) to $N{=}100$.

A layer of beads was fixed at the lateral and bottom walls of a cylindrical container of 
height $20d$ to provide rough boundaries. The inner diameter of the container was $29d$ 
and a rotating four-blade vane of diameter $9d$ was constructed by touching rigid beads. 
The container was first loaded with chains of equal size and relaxed into equilibrium 
under gravity. Next, the vane was rotated in an oscillatory fashion with $1\text{Hz}$ 
frequency and $\pi{/}50\,\text{rad}$ amplitude for 240 s. The oscillatory shear tests 
were carried out at $0.1\text{Hz}$ frequency and for different amplitudes. The total 
torque exerted on the vane was measured after each $\Delta\phi{=}10^{-3}\text{rad}$ 
change in the deflection angle.}

\bibliography{Refs}

\smallskip\smallskip\footnotesize{
\noindent\textbf{Acknowledgments} \\
M.H.\ acknowledges funding from the Netherlands Organization for Scientific Research 
through NWO-VIDI grant No.\ 680-47-548/983. M.R.S.\ acknowledges support by the Deutsche 
Forschungsgemeinschaft (DFG) through Collaborative Research Center SFB 1027 and by the 
Young Investigator Grant of the Saarland University, Grant No.\ 7410110401. 
\smallskip

\noindent\textbf{Author contributions} \\
M.H. designed the research; M.H.\ and G.G.-R.\ planned the experiments; G.G.-R.\ and 
S.Z.\ carried out the experiments; G.G.-R., S.Z., M.H., and M.R.S.\ analyzed the data; 
All authors contributed to the interpretation of the results; M.R.S.\ designed and 
performed the simulations and wrote the manuscript. M.R.S.\ and G.G.-R.\ equally 
contributed to this work. Correspondence and requests for materials should be 
addressed to M.\,H.\ (mehdi.habibi@wur.nl) or M.\,R.\,S.\ (shaebani@lusi.uni-sb.de). 
\smallskip

\noindent\textbf{Competing interests} \\
The authors declare that they have no competing interests. 
\smallskip

\noindent\textbf{Data and materials availability} \\
All data needed to evaluate the conclusions in the paper are present in the 
paper. Additional data related to this paper may be requested from the authors.
\smallskip

\noindent\textbf{Supplementary materials}\\
\noindent Movie S1.\ Examples of oscillatory shear experiments with chain length 
$N{=}\,1$ or $9$ and rotation amplitude $\Phic\,{=}\,0.063$ or $0.632\,\text{rad}$.} 

\end{document}